\documentclass[twocolumn,prb,amssymb,amsmath,superscriptaddress,showpacs]{revtex4}
\usepackage{graphicx}

\begin{document}

\title{Interference induced metallic-like behavior of a two-dimensional
hole gas in asymmetric GaAs/In$_{x}$Ga$_{1-x}$As/GaAs quantum well }

\author{G.~M.~Minkov}
\affiliation{Institute of Metal Physics RAS, 620219 Ekaterinburg,
Russia}
\author{A.~V.~Germanenko}
\author{O.~E.~Rut}
\affiliation{Institute of Physics and Applied Mathematics, Ural
State University, 620083 Ekaterinburg, Russia}

\author{A.~A.~Sherstobitov}
\affiliation{Institute of Metal Physics RAS, 620219 Ekaterinburg,
Russia}
\author{B.~N.~Zvonkov}
\affiliation{Physical-Technical Research Institute, University of
Nizhni Novgorod, 603600 Nizhni Novgorod, Russia}

\date{\today}

\begin{abstract}
The temperature and magnetic field dependences of the conductivity
of the heterostructures with asymmetric  In$_x$Ga$_{1-x}$As quantum
well are studied. It is shown that the metallic-like temperature
dependence of the conductivity observed in the structures
investigated is quantitatively understandable within the whole
temperature range, $T=0.4-20$~K. It is caused by the interference
quantum correction at fast spin relaxation for  $0.4$~K$ < T <
1.5$~K. At higher temperatures, $1.5$~K$<T<4$~K, it is due to the
interaction quantum correction. Finally, at $T>4-6$~K, the
metallic-like behavior is determined by the phonon scattering.
\end{abstract}


\maketitle

Transport properties  of two dimensional (2D) systems reveal the
intriguing features. One of them is a metallic-like temperature
dependence of the conductivity, $\sigma$, at low temperature
$d\sigma/dT<0$. As a rule, such a behavior is observed in the
structures with strong hole-hole ({\em h-h}) or electron-electron
({\em e-e}) interaction characterized  by large value of the gas
parameter $r_s=\sqrt{2}/(a_B k_F)\gg 1$, where $a_B$ and $k_F$ are
the effective Bohr radius and the Fermi quasimomentum, respectively.
That is why the {\em e-e} ({\em h-h}) interaction is considered as
the main reason for the metallic-like $T$-dependence of
$\sigma$.\cite{Pud} The theory of the interaction quantum
correction\cite{Zala01,Gor03} developed beyond the diffusive regime
is widely applied to treat the experimental
results.\cite{Prosk02,Pud03}

On the other hand, there is a  mechanism, which could in principle
result in the metallic-like behavior of the system of weakly
interacting electrons (holes).\cite{Skv98,Geller98,Gor98,Golub02} It
is the interference quantum correction to the
conductivity:\cite{Hik80}
\begin{eqnarray}
\delta\sigma^{\text{WaL}}
&=&G_0\left\{-\frac{1}{2}\ln\left(\frac{\tau}{\tau_\phi}\right)+\ln\left(\tau
\left[\frac{1}{\tau_\phi}+\frac{1}{\tau_s}\right]\right)\right.\nonumber \\
&+&\left.\frac{1}{2}
\ln\left(\tau\left[\frac{1}{\tau_\phi}+\frac{2}{\tau_s}\right]\right)\right\},\,\,\,
G_0=\frac{e^2}{2\pi^2 \hbar},
 \label{eq05}
\end{eqnarray}
where $\tau_s$, $\tau_\phi$ and $\tau$ are the spin, phase, and
transport relaxation time, respectively. The sign of $d\sigma/d T$
for this mechanism depends on the relation between $\tau_s$ and
$\tau_\phi$. When $\tau_s$ is appreciably shorter than $\tau_\phi$
the interference correction is
\begin{equation}
 \delta\sigma^{\text{WaL}}\simeq \frac{1}{2}\,
 G_0\ln{\frac{\tau_\phi}{\tau}}.
 \label{eq10}
\end{equation}
Since the phase relaxation time at low temperatures is determined by
electron-electron collisions, it is inversely proportional to the
temperature, and, thus, the sign of $d\sigma/d T$ given by
Eq.~(\ref{eq10}) is negative. In contrast to the weak localization
characterized by the positive sign of $d\sigma/d T $, this
phenomenon is known as the weak {\em anti}localization (WaL). If the
interference correction dominates the interaction correction, such a
2D system should demonstrate the metallic-like behavior.

\begin{figure}
\includegraphics[width=0.8\linewidth,clip=true]{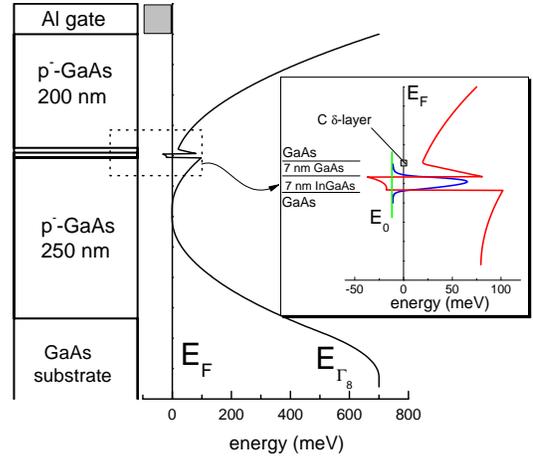}
  \caption{(Color online)
The cross section and calculated valence band profile for the
structure investigated at $V_g=0$~V. }\label{germ0}
\end{figure}

In this paper,  we present results of the experimental study of the
transport in the hole single quantum well with the asymmetric
doping.  Such a manner of doping results in the relatively large
value of the spin-orbit splitting of the energy spectrum due to the
Bychkov-Rashba mechanism\cite{Bych84} and, as consequence, in the
relatively fast spin relaxation. We show experimentally that such a
system really demonstrates the metallic-like behavior caused by the
spin relaxation.

The heterostructure studied was grown by metal-organic vapor-phase
epitaxy on a semi-insulator GaAs substrate. It  consists of a
$250$~nm-thick undoped GaAs buffer layer,  a $7$~nm
In$_{0.2}$Ga$_{0.8}$As well, a $7$~nm spacer of undoped GaAs, a
carbon $\delta$-layer   and $200$~nm cap layer of undoped GaAs. Two
samples from the same wafer have been studied. The samples were mesa
etched into standard Hall bars. In order to tune the hole density,
an Al gate electrode was deposited onto the cap layer. The hole
density and mobility  are $p=(4.0-7.5)\times 10^{11}$~cm$^{-2}$ and
$\mu=(4000-8000)$~cm$^2$/V~s respectively. The valence band profile
calculated  self-consistently is shown in Fig.~\ref{germ0}. Similar
results were obtained for both samples. Here we present the data
obtained for one of them which  parameters for some gate voltages
are presented in Table~\ref{tab1}.

\begin{table}[b]
\caption{The parameters of structure for different gate voltages}
\label{tab1}
\begin{ruledtabular}
\begin{tabular}{cccccccc}
$V_g$ (V) & $p$ (cm$^{-2}$) &$\tau$ (ps) & $\tau_\phi$ (ps) & $\tau_s$ (ps) & $F_0^\sigma$ & $\widetilde{F}_0^\sigma$\\
 \colrule
 0          &$6.5\times 10^{11}$   &$0.73$   &$28/T$  & 4.8 & -0.39 & -0.34\\
 1.3          &$4.8\times  10^{11}$   &$0.48$   &$20/T$ & 11.0 & -0.405 & -0.36\\
 1.6 & $4.4\times  10^{11}$ & $0.39$ & $27/T$ & $17$ & $-0.378$ & $-0.322$\\
 \end{tabular}
\end{ruledtabular}
\end{table}

\begin{figure}
\includegraphics[width=0.8\linewidth,clip=true]{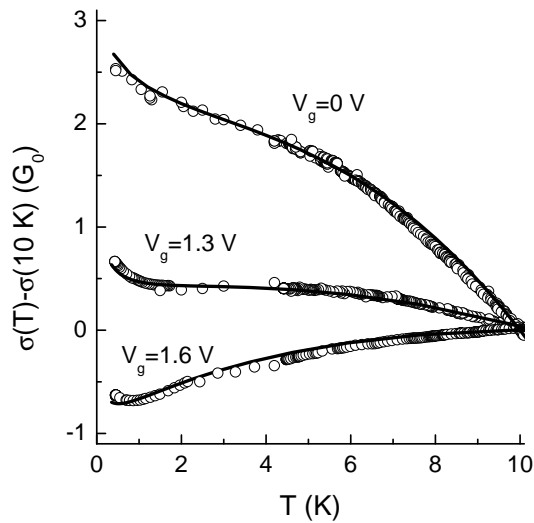}
  \caption{The temperature dependence of the conductivity measured
  at different gate voltages. Symbols are the data,
  lines are the calculation results.}\label{germ1}
\end{figure}

\begin{figure}
\includegraphics[width=\linewidth]{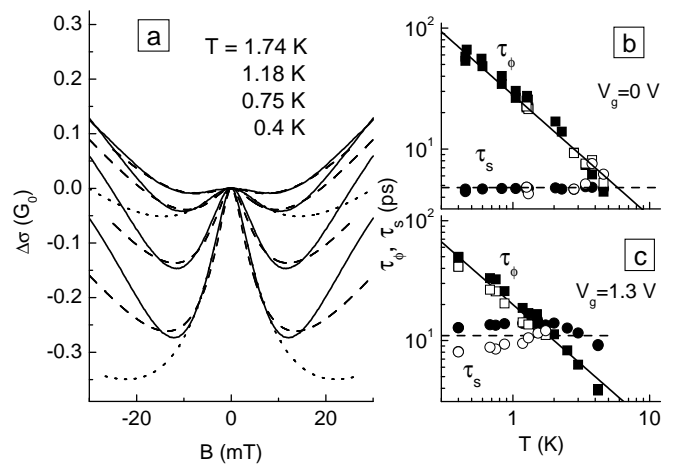}
\caption{(a) -- The $\Delta\sigma$-vs-$B$ dependence for
$V_g=1.3$~V.  Solid lines are experimental data. Dashed lines are
the results of the fitting procedure performed at $B<0.2\,B_{tr}$
(dashed lines) and $B<0.1\,B_{tr}$ (dotted lines). (b) and (c) --
The $T$-dependence of $\tau_\phi$ and $\tau_s$ obtained from the fit
of the magnetoconductivity curves for $V_g=0$~V and 1.3~V,
respectively at $B<0.2\,B_{tr}$ (solid symbols) and $B<0.1\,B_{tr}$
(open symbols), $B_{tr}=91$~mT. Lines are the interpolating
dependences which are used in calculation of
$\delta\sigma^{\text{WaL}}$ and $\sigma(T)$ presented in
Figs.~\ref{germ1} and ~\ref{germ3}.}\label{germ2}
\end{figure}

The temperature dependences of the conductivity measured for
different gate voltages in the absence of a magnetic field are
presented in Fig.~\ref{germ1}. It is seen that they are
metallic-like at the high hole density and mostly insulating at
lower one.  In order to understand the origin of such a behavior,
the magnetic field dependences of the resistivity and the Hall
effect has been investigated. The magnetoconductivity
$\Delta\sigma(B)=\rho_{xx}^{-1}(B)-\rho_{xx}^{-1}(0)$  measured  for
one of the gate voltage, $V_g=1.3$~V, at different temperatures are
shown in Fig.~\ref{germ2}(a). A characteristic antilocalization
minimum is clearly seen; the lower is the temperature, the deeper is
the minimum. As shown in Ref.~\onlinecite{Min05}, the Dyakonov-Perel
mechanism\cite{Dyak71} is the main mechanism of spin relaxation in
such a type systems. It has been also shown that the Bychkov-Rashba
mechanism is responsible for the spin-orbit splitting of the energy
spectrum in $p$-type GaAs/In$_x$Ga$_{1-x}$As/GaAs quantum wells, and
this splitting  is cubic in quasimomentum. In this case the spin
relaxation rate can be obtained  from the fit of the
magnetoconductivity curve by the Hikami-Larkin-Nagaoka
expression.\cite{Hik80} The results of the best fit are shown in
Fig.~\ref{germ2}(a). Since the theory\cite{Hik80} was developed
within the diffusion approximation, the fit was performed for the
relatively low magnetic fields, $B<0.1\,B_{\text{tr}}$ and
$B<0.2\,B_{\text{tr}}$, where $B_{\text{tr}}=\hbar/(2el^2)$ is the
transport magnetic field, $l$ is the transport mean free path.

The parameters $\tau_s$ and $\tau_\phi$ found from the fit at
different temperatures  for $V_g=0$~V and $1.3$~V are shown in
Figs.~\ref{germ2}(b) and \ref{germ2}(c), respectively. For
$V_g=0$~V, the data obtained from the different fitting range of
magnetic field practically coincide. For $V_g=1.3$~V, they are
somewhat different that indicates an accuracy in determination of
the phase and spin relaxation times. One can see that both
$\tau_\phi$ and $\tau_s$ demonstrate reasonable behavior. The
$\tau_\phi$~vs~$T$ dependence is close to $T^{-1}$-law that is
typical for the 2D systems. Within the limits of experimental error
the spin relaxation time $\tau_s$ is mostly independent of the
temperature as should be in the degenerated  gas of carriers.

\begin{figure}
\includegraphics[width=\linewidth, clip=true]{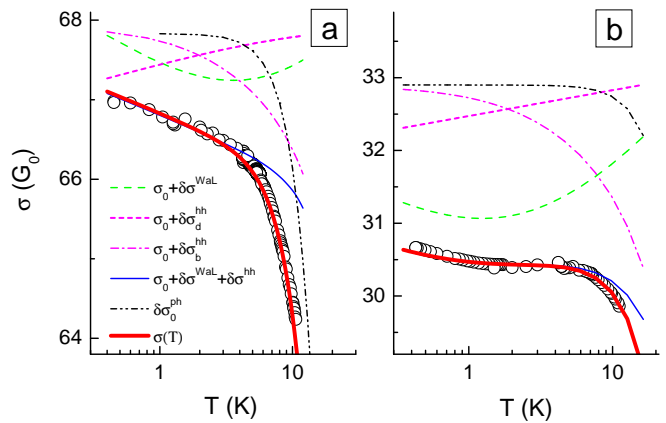}
\caption{(Color online) The $T$-dependence of $\sigma$ for $V_g=0$~V
(a) and $1.3$~V (b). Symbols are the experimental results. Lines are
the results of calculation. }\label{germ3}
\end{figure}

The value of $\tau_s$ is significantly less than that of $\tau_\phi$
only at $T\lesssim (1-3)$~K depending on the gate voltage, that
should provide the metallic-like temperature dependence of the
conductivity only at the lowest temperatures. This is justified by
the calculation results presented in Fig.~\ref{germ3}. The dashed
lines in this figure show the temperature dependence of the
conductivity calculated as
$\sigma(T)=\sigma_0+\delta\sigma^{\text{WaL}}(T)$, where
$\sigma_0=e^2\tau\, n /m$ is the Drude conductivity determined in
the given case by the ionized impurity scattering.\cite{fnt1}  The
dependences $\delta\sigma^{\text{WaL}}(T)$ have been calculated from
Eq.~(\ref{eq05}) with $\tau_s$ and $\tau_\phi(T)$ presented in
Table~\ref{tab1} that well interpolate the corresponding
experimental data shown in Figs.~\ref{germ2}(b) and \ref{germ2}(c).
Thus, the weak antilocalization explain the negative sign of
$d\sigma/d\,T$ only at lowest temperatures.

A different mechanism which can contribute to the temperature
dependence of the conductivity is the {\em h-h} interaction. It can
be conventionally subdivided into the ballistic and diffusion parts
$\delta\sigma_b^{\text{hh}}$ and $\delta\sigma_d^{\text{hh}}$,
respectively:\cite{Zala01,Min05-1}
\begin{subequations}
\label{eq20}
\begin{eqnarray}
\delta\sigma^{\text{hh}}(T)&=&\delta\sigma_b^{\text{hh}}(T)+\delta\sigma_d^{\text{hh}}(T),\label{eq20a} \\
 \delta \sigma_{b}^{\text{hh}}(T)&=&2\pi G_0
\frac{T\tau}{\hbar}\left[1-\frac{3}{8}f(T\tau)\right. \label{eq20b} \\
&&\left.+\frac{3\widetilde{F}_0^\sigma}{1+\widetilde{F}_0^\sigma}
\left(1-\frac{3}{8}t(T\tau,\widetilde{F}_0^\sigma)\right)\right],
\nonumber \\
 \delta \sigma_{d}^{\text{hh}}(T)&=&
 -G_0\left[1+3\left(1-\frac{\ln(1+F_0^\sigma)}{F_0^\sigma}\right)\right]\ln{\frac{\hbar}{T\tau}}, \label{eq20c}
\end{eqnarray}
\end{subequations}
where the functions $f(T\tau)$ and $t(T\tau,\widetilde{F}_0^\sigma)$
are given in Ref.~\onlinecite{Zala01}.  The values of the
Fermi-liquid constants $F_0^\sigma$ and $\widetilde{F}_0^\sigma$
(see Table~\ref{tab1}) have been found as described in
Ref.~\onlinecite{Min05-1}, where the interaction correction was
thoroughly studied in analogous systems.

Dotted and dash-dotted curves in Fig.~\ref{germ3} are drown with
taking into account only the diffusion and ballistics parts of the
interaction correction, respectively. It is seen that their
contributions to the temperature dependence are of different sign:
$\delta \sigma_{b}^{\text{hh}}$ reveals the metallic like behavior
whereas $\delta \sigma_{d}^{\text{hh}}$ is insulating.  Thin solid
lines in Fig.~\ref{germ3} are the total conductivity calculated as
$\sigma(T)=\sigma_0+\delta\sigma^{\text{WaL}}(T)+\delta\sigma^{\text{hh}}(T)$.
It is seen that the consideration of both quantum corrections allows
us to describe  the experimental data in wider temperature range
quantitatively. A good agreement is already evident up to $T\simeq
4-6$~K.

Finally,  the decrease of the conductivity at higher temperature,
$T\gtrsim 4-6$~K, is simply due to the phonon scattering, which
cannot be ignored in heterostructures based on the piezoelectric
crystals as our analysis shows. The dashed-double-dotted lines in
Fig.~\ref{germ3} are the theoretical $T$-dependence of the Drude
conductivity calculated taking into account the phonon scattering
$\sigma_0^{ph}=\frac{e^2n}{m}\left(\tau^{-1}+\tau_{ph}^{-1}\right)^{-1}$,
where $\tau_{ph}^{-1}$ is the phonon contribution to the scattering
rate. The latter is calculated in Ref.~\onlinecite{Karpus90} and
represented in the form
$\tau_{ph}(T)=a(T_0/T)^3\left[1+b(T/T_0)^2\right]$, where
$T_0=\sqrt{2mE_FS_t^2}$ with $E_F$ and $S_t$ as the Fermi energy and
transverse sound velocity, respectively. The parameters $a$ and $b$
depend on the carriers parameters and crystal properties, and have
been calculated according to Ref.~\onlinecite{Karpus90} with
$m=0.16\, m_0$.\cite{Min05-1} One can see that $\sigma_0^{ph}(T)$
really exhibits a considerably strong decrease explaining, thus, the
experimental dependence $\sigma(T)$ at $T\gtrsim 4-6$~K.

The bold lines in Fig.~\ref{germ3} are calculated taking into
account all three effects considered above: the interference quantum
correction, the hole-hole interaction quantum correction, and the
temperature dependence of the Drude conductivity due to phonon
scattering:
\begin{equation}
 \sigma(T)=\sigma_0^{ph}(T)+\delta\sigma^{\text{WaL}}(T)+\delta\sigma^{\text{hh}}(T).
 \label{eq40}
\end{equation}
It is clearly seen that the calculation results are in excellent
agreement with the experimental ones (see also Fig.~\ref{germ1}).

Let us briefly discuss what type of the temperature dependence of
the conductivity is expected at millikelvin temperatures. It is
governed by competition between the interference and interaction
quantum corrections. The interference correction
$\delta\sigma^{\text{WaL}}$ at low temperatures has the form given
by Eq.~(\ref{eq10}) and provides the logarithmic metallic-like
behavior of $\sigma(T)$.

The diffusion part of the interaction correction
$\delta\sigma^{\text{hh}}_d$ is logarithmic as well as
$\delta\sigma^{\text{WaL}}$,  but cannot be, however, described by
Eq.~(\ref{eq20c}). The fact is that the second term in square
brackets in Eq.~(\ref{eq20c}) is suppressed at low temperature,
$T\ll \hbar/\tau_s$, so that the expression for
$\delta\sigma^{\text{hh}}_d$ consists of only the Fock
term\cite{Gor98,AA83}
\begin{equation}
\delta\sigma^{\text{hh}}(T)\simeq
\delta\sigma_d^{\text{hh}}(T)\simeq
 -G_0\ln{\frac{\hbar}{T\tau}},\,\,\,\, T\ll \hbar/\tau_s.
 \label{eq25}
\end{equation}

In contrast to $\delta\sigma^{\text{hh}}_d$, the ballistic
contribution of the interaction correction
$\delta\sigma^{\text{hh}}_b$ is insensitive to the Bychkov-Rashba
spin-relaxation mechanism,\cite{Dmit07} and vanishes at $T\to 0$ in
accordance with  Eq.~(\ref{eq20}). This is because the relevant
ballistic paths contain two segments with opposite momenta, the spin
rotation on the first segment is exactly canceled by
counter-rotation of spin on the second segment.

Thus  the total quantum correction
$\delta\sigma^{\text{WaL}}+\delta\sigma^{\text{hh}}$ should be
negative at very low temperatures, and the $T$-dependence of the
conductivity is anticipated to be insulating-like
\begin{equation}
 \sigma(T)\propto \delta\sigma^{\text{WaL}}(T)+\delta\sigma^{\text{hh}}(T)= -\frac{1}{2}\,G_0\ln{\frac{\hbar}{T\tau}}.
 \label{eq30}
\end{equation}
Using the  values of $\tau_s$ from Table~\ref{tab1} we estimate that
this regime should happen in our samples at $T\ll 0.4-1.5$~K. The
experiments intended to verify this prediction are in our plan for
the future.

In conclusion,  we have experimentally studied the temperature
dependence of the conductivity of 2D hole gas in asymmetrical
GaAs/In$_{x}$Ga$_{1-x}$As/GaAs quantum well heterostructures. It is
shown that there is not one universal mechanism which is responsible
for the metallic-like temperature dependence of the conductivity in
the whole temperature range for the structures investigated. There
are three physically  different mechanisms: the weak
antilocalization, the hole-hole interaction, and the phonon
scattering. Each of them is the main within the corresponding
temperature range. The first one dominates at $T\simeq 0.4-1.5$~K.
At higher temperatures, $T\simeq 1.5-4$~K, the interaction quantum
correction becomes essential. Finally,  the phonon scattering
becomes  important at $T\gtrsim 4-6$~K.

We are grateful to I.~V. Gornyi for very useful discussions. This
work was supported in part by the RFBR (Grant Nos. 05-02-16413,
06-02-16292, and 07-02-00528), the CRDF (Grant No. Y3-P-05-16), and
by a Grant from the  President of Russian Federation for Young
Scientists.

\end{document}